\documentclass[aps, preprint, a4paper, notitlepage]{revtex4-1}
\usepackage[utf8]{inputenc}
\usepackage{amsmath}
\usepackage{amsfonts}
\usepackage{amssymb}
\usepackage{graphicx}
\usepackage{siunitx}
\usepackage{xcolor}

\begin{document}
\title{Collective motion of granular matter subjected to swirling excitation}
\author{Song-Chuan Zhao}
\affiliation{State Key Laboratory for Strength and Vibration of Mechanical Structures, School of Aerospace Engineering, Xi’an Jiaotong University, Xi’an 710049, China}
\affiliation{Institute for Multiscale Simulation, Friedrich-Alexander-Universit\"{a}t, Cauerstra\ss e 3, 91058 Erlangen,Germany}
\author{Thorsten P\"{o}schel}
\affiliation{Institute for Multiscale Simulation, Friedrich-Alexander-Universit\"{a}t, Cauerstra\ss e 3, 91058 Erlangen,Germany}

\begin{abstract}
A two-dimensional granular packing under horizontally circular shaking exhibits various collective motion modes depending on the strength of the oscillation and the global packing density. For intermediate packing density and oscillation amplitude, a high density phase travels along the container's side wall in clockwise direction, while the oscillation itself is anti-clockwise.
Further increasing packing density towards the hexagonal packing, the whole packing rotates collectively in clockwise direction. The core of the packing rotates as a solid and is separated from the boundary by a fluid-like layer. Both motion modes are associated with the asymmetric motion of particles close to the side wall.  
\end{abstract}
\maketitle

%\section{introduction}
Collective motion is observed in a wide range of many-body systems at all scales far from equilibrium,  ranging from mammal herbs and traffic jams down to bacteria colonies and the cooperative action of molecular motors. 
%Though the interaction detail is very different in these system, the displayed collective behavior produces similar patterns of extended spatiotemporal coherence, which can be seen as the result of spontaneous symmetry-breaking~\cite{Redner2016}. 
The universal features, which can be seen as the result of spontaneous symmetry-breaking~\cite{Redner2016}, stimulate the study of the general underlying principles of collective motion. %The first theoretical effort is to establish a minimal model merely including the necessary principles. 
A frequently used theoretical approach is the Vicsek model and its variants~\cite{vicsek1995,chate2008,solon2013}, where the system is simplified as an ensemble of active particles with short-ranged alignment interaction. Experimental comparison in such a simple scenario is desirable.
However, it is not easy to preform experiments with biological systems, as the activity of and the interaction between individual particles are often not well controlled. 
On the other hand, vibrating granular matter exhibits spontaneous symmetry-breaking and inspired experimental studies of collective behavior~\cite{Aranson2006}. In those experiments, two different types of collective behavior are observed: density patterns, such as clustering~\cite{Kudrolli2008,Narayan2007,Berardi2010} and the directional motion of a persistent cluster~\cite{Deseigne2010,Scholz2018}. The former is sometimes referred to as phase separation. Because of the short-ranged interaction, the spontaneous formation of the dense phase is necessary to the collective motion, but not sufficient. 
%which is observed in biological systems as well and is reported in the Vicsek model. 
In addition, the key ingredient in the Vicsek model, an interaction term aligning the motion of particles must present. 
This distinction is reported in numerical simulations~\cite{Fily2012} and experiments on active colloidal suspensions~\cite{Buttinoni2013,Redner2013}. There, isotropic particles with isotropic interaction produce dynamic clustering without alignment order and collective motion, called motility-induced phase separation~\cite{Tailleur2008, Redner2016}.
Recent research further highlights the possible hidden alignment interactions in isotropic systems, and
 collective motion may occur~\cite{Deseigne2010,Caprini2020}. The alignment interaction there is induced by the interplay between repulsion/collisions of particles and self-propulsion. As a result, particles in the cluster tend to follow their local average velocity. 

Our previous work reports dynamic clustering in a submonolayer of beads subjected to horizontal oscillations~\cite{Zhao2020}, which is caused by a similar mechanism as mentioned above. There, the breaking of symmetry is triggered by the counteraction between the frictional driving from the subtrate and collisions/contacts. In consequence, the particle speed depends on the density of its neighborhood. In such a system, other self-organized, collective motion modes exist~\cite{Kumar2015}.
In this paper, we first investigate the motion mode where a dense phase travels along the boundary of the container, causing a coherent transport of mass. Moreover, the collective motion in the situation where the whole packing appears as a single cluster (2D hexagonal packing) is studied.

\begin{figure*}
	\includegraphics[width=17cm]{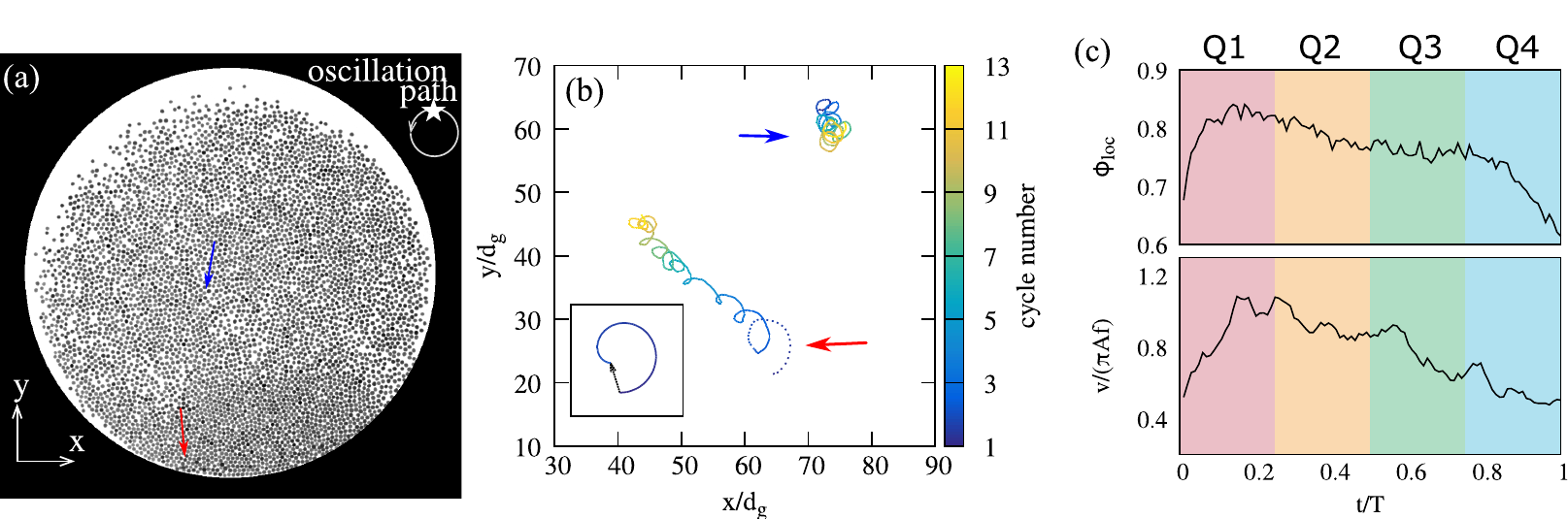}
	\caption{\label{f.boundary_theory}(a) A snapshot of the system in the middle of cycle 1. See SI video for its motion. (b) The trajectories of two particles, whose initial locations are indicated in (a), over the period from cycle 1 to cycle 13. The dashed line highlights the motion in the first cycle. Inset: a trajectory given by Eq.~\ref{eq.traj}. (c) The speed $v$ of the peripheral particle (indicated by the red arrow in (a)) and the local density of its neighborhood $\phi_{loc}$ during the first cycle. The block background correspond to the four quarters, Q1 to Q4.}
\end{figure*}

%\section{experiments}
The system studied here consists of a layer of spherical beads within a cylindrical container that is subjected to anti-clockwise circular oscillation in the horizontal plane. The oscillation is of frequency $f=5\si{Hz}$ and amplitude $A\in [2\dots 13]\si{mm}$. Note that here the container does not rotate , but move in an orbit of a diameter $A$ [cf. Fig.~\ref{f.boundary_theory}a inset]. 
The oscillation can be represented as the superimpose of periodic motion in $x$ and $y$ directions:  $x=(A/2)\sin(2\pi ft)$ and $y=-(A/2)\cos(2\pi ft)$ respectively. The monolayer consists of $N_{tot}$ polydisperse Zirconium Oxide beads of diameter 0.8\dots 1 mm with mean $d_g=0.9\si{mm}$. The grains located on an acrylic plate are confined by a 3D-printed circular side wall of diameter $D=82\si{mm}$. The global packing density is $\Phi=N_{tot}(d_g/D)^2$. The system is leveled carefully such that the inclination of the bottom plate is smaller than 0.02 mm/m. The packing is illuminated by a LED panel from below, and the dynamics are captured by a high speed camera (Mikrotron MC1362) at the top, The camera is fixed in the laboratory frame of reference, recording at a frame rate of 500 \si{Hz}.

The system displays various collective motion modes depending on $\Phi$ and $A$. We define two characteristic packing densities, $\Phi_{hex}=\sqrt{3}\pi/6$ and $\Phi_{min}=0.51$. The former corresponds to the hexagonal structure that is the closest packing in two dimensions. For $\Phi<\Phi_{min}$ no persistent collective motion is observed in the range of $A$ explored. Above $\Phi_{min}$, the packing exhibits spontaneous clustering near the center for $A$ exceeding a threshold $A_c$~\cite{Zhao2020}. Meanwhile, unique motion modes appear below $A_c$, which are the focus of this paper.

We first investigate a system at $\Phi=0.57$. The system is largely uniform for $A<7\si{mm}$.  
The particles are driven by the frictional force arsing from the motion of the substrate, which not only accelerates the linear momentum of particles but also their angular momentum. In the steady state, there is no relative motion at the contacting point between the particle and the bottom plate, and the linear speed of the particles reaches 2/7 of that of the oscillation due to the rotational degree of freedom~\cite{Kondic1999,Chung2011}. ON one hand, this velocity difference results into a region of no particles (see Fig.~\ref{f.boundary_theory}a), and the effective packing density is $\Phi_0=\Phi/\alpha$. A good approximation gives $\alpha=(1-5A/7D)$.\cite{Zhao2020}
%Therefore, individual particles move in a circular fashion, resembling the oscillation path qualitatively. 
On the other hand, this implies that a particle would move faster, if its rotational degree of freedom is suppressed. This counteraction between the linear and angular momentum transfer is crucial to spontaneous clustering when $A>A_c$~\cite{Zhao2020} and pattern formation of granular matter under 1D horizontal shaking~\cite{Krengel2013}. It also plays a key role in the motion modes studied here, which will be discussed later.

When increasing the amplitude, at $A= 7 \si{mm}$ a region of high density appears near {the side wall} of the container, see Fig.~\ref{f.boundary_theory}a. Consider now particles less than $5/7A$ away from the side wall. In the first half of the oscillation cycle, they collide with the side wall, and a compression shock front is formed which leaves the area behind it compacted to $\sim\Phi_{hex}$. The incompressibility of the granular packing close to $\Phi_{hex}$ enhances the propagation of the shock towards the center of the packing. According to this simple picture, the width of the condensed area is about $(5/7)A\Phi_0/(\Phi_{hex}-\Phi_0)\approx 11\si{mm}$ here, in good agreement with the observation in Fig.~\ref{f.boundary_theory}a. In the second half of the oscillation cycle, the condensed area gradually dilutes (see SI video). The same process is repeated the next cycle. However, this condensed phase is not stationary in space. As can be seen in the SI video, the condensed state travels along the periphery of the container in the clockwise direction, opposite to the oscillation. It takes 13 cycles for the dense phase to complete one circle around the side wall.

We track a typical particle in the {condensed area} at the beginning of the observation period, whose location is indicated by the red arrow in Fig.~\ref{f.boundary_theory}a. This particle is referred to as \textit{peripheral particle} in the following. Its trajectory is plotted in Fig.~\ref{f.boundary_theory}b. Within 13 cycles the peripheral particle moves a distance of $35d_g$, in comparison with the length of the {side wall} $286d_g$, over which the {condensed phase} travels during the same period. This indicates that the observed motion is the propagation of the density pattern rather than the motion of a persistent cluster of particles. The travel speed of the density pattern is significantly faster than that of the particles. 
In fact, the particle is left behind by the {condensed phase} after the fourth cycle. Its trajectory is then changed from spirals to more circular-like. The trajectory of a particle in the central area (indicated by the blue arrow in Fig.~\ref{f.boundary_theory}a and b) is plotted for comparison. This central particle is never reached by the shock. It does not show significant displacement and moves in a circular manner. Its average speed is $0.3\pi Af$, representing the expected ratio of $2/7$ with respect to the oscillation speed $\pi Af$. 
 It is thus obvious that the propagation of the dense phase is caused by the coherent spiral motion of particles near the circumference. The question about the origin of the spiral trajectories is raised. It is insightful to closely investigate the motion of the peripheral particle.

The speed of and the local density around the peripheral particle during the first cycle are plotted in Fig.~\ref{f.boundary_theory}b. The local density $\phi_{loc}$ is computed for individual particles in a circular neighborhood with a diameter of $5d_g$~\footnote{Using a neighborhood size of $7d_g$ does not change the conclusion qualitatively.}. It can be seen that $\phi_{loc}$ and the speed $v$ follow a similar evolution pattern: increase dramatically in the beginning, saturate, then decay. We divide the oscillation cycle into quarters Q1 to Q4, as shown in Fig.~\ref{f.boundary_theory}c. In Q1 and Q2 the container moves from the minimum $y$ towards the maximum. Particles close to the lower side wall are compressed together and forced to move along with the container, obtaining a higher speed than those outside of the {condensed area}. This concludes the first half of the cycle, the compression process. In the second half of the oscillation (Q3 and Q4) the {container} reverses its motion in $y$ direction. Without the confinement of the {side wall} the {condensed area} gradually dilates, and $v$ decays. If there are no interactions between particles, $v$ would have quickly decreased to $(2/7)A\pi f$. However, the frictional interaction between particles effectively suppresses the rotational degree of freedom of particles during contacts~\cite{Krengel2013}, and a dense neighborhood enhances this effect~\cite{Zhao2020}. According to the counteraction between the linear and angular momentum transfer, this results into a dependence of $v$ on $\phi_{loc}$~\footnote{As it is observed in Ref.~\cite{Zhao2020}, the dependence of $v$ on $\phi_{loc}$ is linear in the region of $0.6<\phi_{loc}<0.75$.}, when the confinement of the {side wall} is absent. Therefore, the decay of $v$ is gradual over Q3 and Q4, at the same pace as $\phi_{loc}$ does.  
 As a result, the compression and dilation processes are asymmetric. After a full oscillation cycle the peripheral particles obtain a net motion in both $x$ and $y$ directions, which promotes the transport of the high density region.

In summary, the observed collective motion in Fig.~\ref{f.boundary_theory} is the result of the breaking  of time reversal symmetry of the particle motion in one oscillation cycle. The resultant trajectory of the  particles is a spiral. Write the velocity components of a particle as a fraction of the oscillation velocity:
\begin{equation}
v_x=C(t) A\pi f\cos\left(2\pi ft\right),\quad v_y=C(t) A\pi f\sin\left(2\pi ft\right),
\label{eq.traj}
\end{equation}
with $C(t)\leq 1$. The asymmetrical motion implies that $C(t)-C(T/2)\neq \pm (C(T-t)-C(T/2))$ in general. Take $C(t)=\left[1-5/7\left(tf\right)^2\right]$ for instance, where it is assumed that particles in the shock follow the oscillation instantaneously then gradually slows down. 
%Assuming the particles in the shock follow the oscillation instantaneously. Assume that $\phi_{loc}$ changes in time, like $1-(tf)^2$, which is the lowest nonlinear order. The velocity in the $x$ and $y$ directions are given by:
%with $C(t)=\left[1-5/7\left(tf\right)^2\right]$. 
The computed trajectory is shown in the inset of Fig.~\ref{f.boundary_theory}b. It qualitatively captures the experimental trajectory in the first cycle (dotted line in the main plot of Fig.~\ref{f.boundary_theory}b). The above analysis is valid not only for the cylindrical container used in the current experiments, but for other convex shapes as well~\cite{Suppl}. 

 For $A<7\si{mm}$ the motion of the condensed boundary layer is not observed. It is plausible that to establish the coherence of particle motion a minimum size of the condensed phase is required. As analyzed above, the size of the shock increases with $A$ and $\Phi$. Therefore, for denser packings (high $\Phi$) the necessary $A$ for triggering this collective motion of the boundary layer decreases. Indeed, for $\Phi=0.76$, the minimum amplitude, where this motion mode is observed, is reduced to $A=4\si{mm}$, and the shock ($\Phi\approx\Phi_{hex}$) occupies half of the system size. How would the system behave, when $\Phi$ is so high that the size of the shock is comparable to the system size? 
% In the next section we consider such a situation.

%\section{High packing density: $\Phi\approx\Phi_{hex}$}

\begin{figure*}
\centering
\includegraphics[width=17cm]{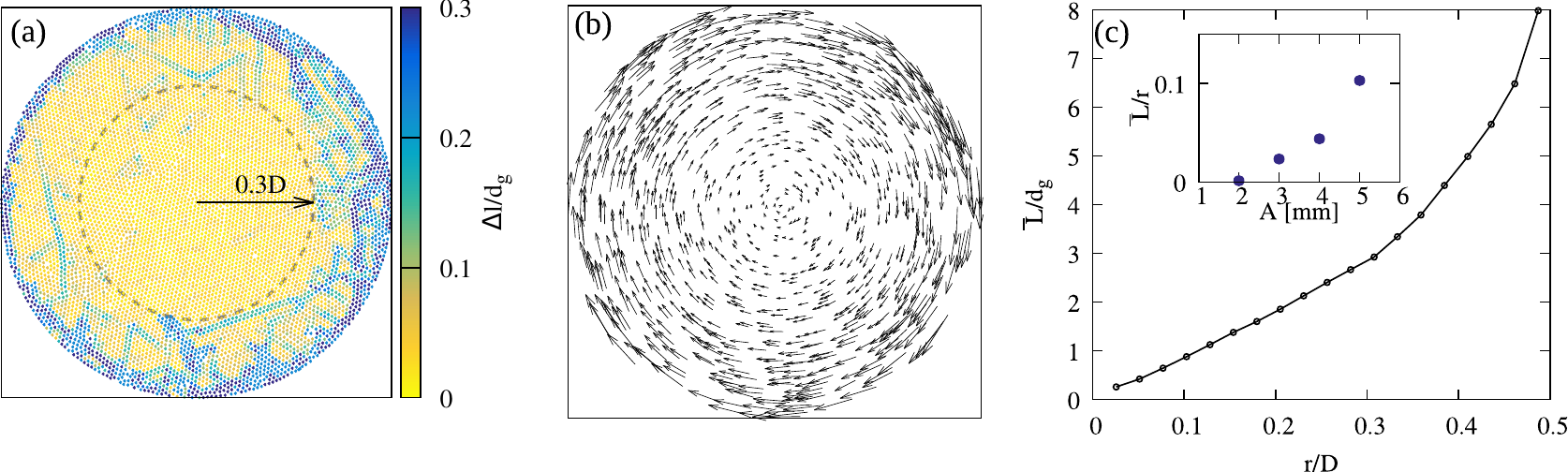}
\caption{\label{f.cryst_motion}(a) A plot of the packing at $\Phi=0.899\approx\Phi_{hex}$. The color denotes the averaged change of neighboring distance, $\Delta l$. See Eq.~\ref{eq.diffuse} and the text for more details. (b) The arrows indicate the traveled distance of individual particles during 30 oscillation cycles. For the ease of visualization, only one tenth of the trajectories is plotted here. (c) The average length of the vectors in (b), $\bar{L}$, as a function of their distance to the center of the packing, $r$. Inset: The averaged $\bar{L}/r$ for $r<0.3D$ for various $A$.}
\end{figure*}

$\Phi=\Phi_{hex}$ marks an extreme in two dimensions. Here, we consider a packing at $\Phi=0.899$. At $A=5\si{mm}$, the packing remains two-dimensional, and the central area of the packing maintains the (poly-)crystalline state with a few defects and dislocations (Fig.~\ref{f.cryst_motion}a). For larger $A$,  particles close to the side wall {jump over} each other, and the two-dimensional scenario breaks down, which raises difficulty of tracking individual particles from the top view. However, the observed motion mode is qualitatively the same as that at lower $\Phi$ examined in Figure~\ref{f.boundary_theory}, \textit{i.e.}, the boundary layer travels along the circumference of the side wall in clockwise direction. In this state, the center becomes dilute. This opens up the possibility of spontaneous clustering via the mechanism explained in Ref.~[\citenum{Zhao2020}], when increasing $A$ up to 11 mm. The last two collective modes ($A>5\si{mm}$) are qualitatively the same to that explored above in Ref.~\citenum{Zhao2020}. Therefore, we instead focus on the low oscillation amplitude experiments here.

In the state of Fig.~\ref{f.cryst_motion}a the whole packing rotates slowly around its center clockwise.~\cite{Suppl} At the first glance, this is similar to the reptation motion reported in~\cite{Scherer1996,Scherer2000}, where the whole packing is regarded as a rigid disk rotating along the {side wall} without sliding. Nevertheless, individual particle tracking reveals differences. The displacement vector of particles over 30 oscillation cycles is plotted in Fig.~\ref{f.cryst_motion}b. The direction of vectors indicates the globally clockwise rotation, and its norm is referred to as $L$. Denote the radial distance of particles to the center of the packing as $r$.  The data of $L$ is binned according to $r$ with a binning size $\Delta r\sim 2d_g$. The average within each bin, $\bar{L}$, is computed and plotted versus $r$ in Fig.~\ref{f.cryst_motion}c. $\bar{L}$ increases with $r$ linearly for $r<0.3D$. The linearity indicates the rigid body rotation of the core of the packing. Beyond this regime, however, the relation becomes nonlinear. In the supplementary video~\cite{Suppl}, one may see that the motion in the region of $r>0.3D$ corresponds to a fluid-like layer. 

To elaborate the fluid-like state, we further compute the averaged change of the distance of individual particles to their neighbors, $\Delta l$. Its definition for particle $i$ is given in Eq.~\ref{eq.diffuse}.
\begin{equation}
\Delta l_i=\frac{1}{6}\sum_{j=1}^6 \left| l_{ij}(30)-l_{ij}(0) \right|
\label{eq.diffuse}
\end{equation}
For particle $i$ its six nearest neighbors $j$ are identified at the beginning of the observation. $l_{ij}(n)$ is the distance between particles $i$ and $j$ at the end of cycle $n$. $n=0$ corresponds the beginning of the observation. $\Delta l$ is a measure of fluidity. In the fluid-like state the neighborhood relation does not persist, resulting into $\Delta l>0$. In contrast, $\Delta l$ is virtually 0 in the solid-like state. The distribution of $\Delta l$ is given in Fig.~\ref{f.cryst_motion}a. The separation of the solid-like core and the fluid-like boundary layer at $0.3D$ is in agreement with that deduced from the relation between $\bar{L}$ and $r$. Note that $\Delta l$ is non-uniform in the region of $r>0.3D$, \textit{e.g.,} there are long-life time crystalline patches in the `glassy' boundary layer. When reducing the oscillation amplitude, the rotation speed of the solid core decreases (see Fig.~\ref{f.cryst_motion}c inset), and the core expanses to $r\approx 0.4D$. At $A=2\si{mm}$ the rotation of the solid core virtually ceases. The separation between the solid core and the fluid layer is stable and reversible with respect to $A$, but is not associated with a first-order like transition as in the vertically vibrated system.~\cite{Pacheco-Vazquez2009}

This global rotation mode can be understood following the same line of argument of the traveling dense boundary layer investigated previously. Because of the curvature of the side wall the crystalline, the packing density is lower than $\Phi_{hex}$ close to the boundary. The particles near the side wall thus experience compression and dilation during each oscillation cycle. Just as discussed above, the asymmetry of the particle motion in the compression and dilation process occurs, introducing the spiral-like motion of particles (Fig.~\ref{f.boundary_theory}b), even though more subtle. This promotes the motion of the fluid-like boundary layer. While this effect is absent in the solid core, the relative motion between the fluid-like layer and the solid-like core incurs a torque. It is plausible that the torque overcomes the friction between the solid-like core and the substrate only for $A$ larger than a threshold, e.g., which is 2mm here.

%\section{Conclusion}
We studied the collective motion of a granular system under swirling excitation. Depending on the packing density $\Phi$ and the oscillation amplitude $A$, two types of collective motion modes are observed. For intermediate $\Phi$ and $A\geq 7\si{mm}$ a condensed phase travels along the side wall of the container, while the center of the packing is unaffected (Fig.~\ref{f.boundary_theory}). For $\Phi$ approaching the closet packing in two-dimension, $\Phi_{hex}$, the whole packing rotates, however, inhomogeneously. A solid-like core is surrounded by a fast moving fluid-like boundary layer (Fig.~\ref{f.cryst_motion}). In both cases, the interplay between the driving and friction forces between particles underlies the coherent transport of mass, which leads to the spiral-like motion of particles near the boundary. 

{Spiral trajectories of particles were found in the reptation mode for the larger $d_g/D$~\cite{Scherer1996,Scherer2000}, though no coexistence of solid and fluid states was reported there. This indicates a potential role played by this aspect ratio. With a similar $d_g/D$ studied here, one dimensional horizontal shaking has been utilized to produce (partially) crystalline three-dimensional sphere packings~\cite{Pouliquen1997}. In contrast to time-variant acceleration under one-dimensional shaking, the orbital oscillation used here exerts accelerations of constant magnitude. It may open up a new protocol of studying the crystallization of athermal particles~\cite{Saadatfar2017}.}

\bibliography{clustering.bib}
\end{document}